# State-dependent Asset Allocation Using Neural Networks

Reza Bradrania[a*] and Davood Pirayesh Neghab[b]

[a]UniSA Business, University of South Australia, Australia
[b]Department of Industrial Engineering, Koc University, Istanbul, Turkey

April 2021

**ABSTRACT:** Changes in market conditions present challenges for investors as they cause performance to deviate from the ranges predicted by long-term averages of means and covariances. The aim of conditional asset allocation strategies is to overcome this issue by adjusting portfolio allocations to hedge changes in the investment opportunity set. This paper proposes a new approach to conditional asset allocation that is based on machine learning; it analyzes historical market states and asset returns and identifies the optimal portfolio choice in a new period when new observations become available. In this approach, we directly relate state variables to portfolio weights, rather than firstly modeling the return distribution and subsequently estimating the portfolio choice. The method captures nonlinearity among the state (predicting) variables and portfolio weights without assuming any particular distribution of returns and other data, without fitting a model with a fixed number of predicting variables to data and without estimating any parameters. The empirical results for a portfolio of stock and bond indices show the proposed approach generates a more efficient outcome compared to traditional methods and is robust in using different objective functions across different sample periods.

*: Corresponding author. Reza.bradrania@unisa.edu.au

We thank Talis Putnins, Gabor Rudolf, who is no longer with us, Adrian Lee, Nigar Hashimzade, Eduardo Roca, Robert Elliott, Ihsan Badshah, Chandra Krishnamurti, and participants at the 2019 UniSA Fintech Conference for their comments and discussions.

## 1. Introduction

The mean-variance framework of Markowitz (1952) has been widely recognized as the foundation of modern portfolio theory. Inspired by this framework, investors traditionally consider a static single-period optimization for their asset allocation problem and optimize efficiency across all economic conditions. The allocation can be revisited regularly, but it is unlikely to change significantly, as long as the aim is efficiency across all market conditions (Nystrup, Madsen and Lindstroms, 2018). However, economic and market conditions change over time and present time-varying investment opportunity sets. It is well documented that the means, variances, covariances and higher order moments of assets are time-varying[1]. If economic conditions are linked to asset returns, a state-dependent asset allocation process should add value over static weights (Harvey and Dahlquist, 2001). In fact, several studies show the importance of state-dependent asset allocation to exploit the predictability of the moments of asset returns and hedge changes in the investment opportunity set in both single-period and multi-period contexts (Barberis, 2000; Ait-Sahalia and Brandt, 2001; Brandt and Clara, 2006; Ang and Bekaert, 2004; Guidolin and Timmermann, 2007; Kritzman, Page and Turkington, 2012, Nystrup, Madsen and Lindstroms, 2018; and Laborda and Olmo, 2017).

The purpose of state-dependent asset allocation strategies is to take advantage of good economic and market conditions (or states) and mitigate the impact of adverse states (Sheikh and Sun, 2012). However, it is very difficult to model a conditional return distribution and find a closed form solution for optimal investment strategies unless strong assumptions on the statistical distribution of asset returns are imposed. The predominant approach includes a two-step process (Laborda and Olmo, 2017); firstly, modeling separate features of the return distribution, and then determining the optimal portfolio choice. However, this approach does not guarantee the optimal allocation for the problem at hand because different sets of state variables may predict different moments of returns used in the objective function (e.g., Ait-Sahalia and Brandt, 2001). Alternative approaches include numerical methods, and stochastic programming (Berberis, 2000; Brennan, Schwartz, and Lagnado, 1997, 1999; Campbell, Chan, and Viceira, 2003 and Campbell, and Viceira, 1999, 2001, 2002) or semi non-parametric methods (Ait-Sahalia and Brandt, 2001 and Brandt and Clara, 2006), which have limitations in the number of and type of state variables used in the process or require fitting a linear model with data. Nevertheless, the dependence of the optimal portfolio choice on the whole return distribution is very complex and far from linear and the predicting power of specific state variables may change over time.

In this paper, we propose a different approach using Artificial Neural Network (ANN), a powerful machine learning tool that uses state variables to directly decide on tactical portfolio weights in the optimization problem. This approach, as a non-parametric method, estimates the optimal portfolio weights for a myopic investor at the beginning of the investment period without fitting a model with a certain number of state variables to the data, without estimating any parameters and without estimating a specific distribution of the data. In this approach, the investor's ultimate objective is to maximize a performance ratio, such as the Sharpe ratio, as a mean-variance utility function[2]. The method uses the historical values of the state variables and returns to train a machine learning network and determine a non-linear model. This model is used at the beginning of the investment period, when the latest values of state variables are available, to estimate the optimal weight for the period ahead.

---

[1] See, e.g., Merton (1980); Fama and French (1988); Barberis (2000); Diebold, Lee, and Weinbach (1994); Chen (2009); Kanas (2008); Costa and Kwon (2019); and Ang and Bekaert (2004).
[2] Performance ratios are considered as the explicit asset allocation formula in which the optimal allocation is proportional to the ratio of the reward to risk. They are used for managing mutual funds and are popular in optimal asset allocations (see, e.g., Farinelli, Ferreira, Rossello, Thoeny, and Tibiletti (2008, 2009) and Biglova, Ortobelli, Rachev and Stoynov (2004)).

In this approach, instead of first modeling moments and various features of the conditional return distribution, and subsequently characterizing optimal trading strategies, we skip the estimation of the conditional return distribution and directly decide on portfolio weights using predicting variables. This is important because the relationship between the portfolio weights – as the complex functions of the return distribution – and the predictors is less noisy than the relationship between the individual moments and the predictors. Therefore, we avoid the potential misspecification and additional noise due to estimation of the return distribution as an intermediary step. Another advantage of our approach is that we do not aim to select specific state variables, as the network of the machine learning disregards irrelevant state variables, and also there is no limitation on the number of state variables used in the optimization problem. This is different from the traditional methods of a conditional portfolio advice in which selection or combination of state variables play an important role in the conditional asset allocation.

We examine the proposed method empirically in an investment universe consisting of the Standard and Poor (S&P) 500 index and the US benchmark thirty-year government bond index and compare its performance to that of three traditional methods. We study the optimal asset allocation for a myopic investor who has a mean-variance utility function, such as the Sharpe ratio, and maximizes for one-period-ahead the trade-off between the desired reward and risk. In addition to Sharpe, we use various performance ratios that are widely used in the literature as the objective function in the portfolio problem, to show the robustness of our proposed procedure. Each performance ratio captures different risk preferences and computes return and risk differently because it incorporates different moments of returns in calculations. More specifically, we use Sharpe, Mean Absolute Deviation (MAD), MiniMax, Gini, CVaR, and Rachev ratios, which include both symmetric and asymmetric ratios[3]. We also employ four popular state variables: the default spread, the dividend yield (of the S&P 500 Index), the term spread, and an S&P 500 Index trend (or momentum). The data covers from January 1986 to the end of December 2018.

As well as our ANN-based approach, we apply three traditional methods, as benchmarks. The first two methods estimate the moments of returns (mean and covariance) using historical data before maximizing the ratios, whereas the third approach integrates the estimation and optimization steps to determine the optimal portfolio weights. In the first benchmark method, a VAR model (which is an autoregressive process with one lag) is used to estimate the moments of returns, whereas in the second benchmark method a factor model is used to estimate moments. The factor model includes the four state variables that we have used in this paper. In the third benchmark method, we use a linear, parametric method based on Brandt et al. (2009) approach that avoids the estimation of the conditional distribution of returns, and directly estimates the relationship between state variables and asset weights. The objective function in this approach is the expected utility that would implicitly take into account the different moments of the return distribution and the risk profile of the investor. The results from the full out-of-sample (1999–2018) as well as sub-sample periods show that all ratios are greater when we use the proposed ANN-based approach to find the optimal weights compared to the traditional methods. For example, across the full sample, the monthly Sharpe ratio (CVaR ratio) is 0.09 (0.34) using our approach, whereas it is 0.06 (0.17), 0.07 (0.20), and 0.08 (0.22) based on the first, second and third benchmarks, respectively. Our findings show that the improvement for all ratios using ANN, compared to the benchmarks, is not only due to considering the state of the market in portfolio choice, but to capturing the nonlinear relations between these variables and the portfolio weights.

---

[3] The Sharpe ratio is valid for the portfolio choice of an investor if returns are normally distributed because mean and variance can explain their distribution; however, it is well documented that returns do not follow normal distributions (Leland, 1999; Kat and Brooks, 2002; Agarwal and Naik, 2004; and Malkiel and Saha, 2005). When returns exhibit heavy tails or present kurtosis or skewness, using Sharpe as the objective function leads to an inefficient estimation of risk, and incorrect investment decisions. To overcome this well-known limitation of the Sharpe ratio, several alternative performance ratios have been proposed in the literature. For example, in Gini (Shalit and Yitzhaki, 1984), Mean Absolute Deviation (MAD) (Konno and Yamazaki, 1991) and Minimax (Young 1998), the risk measures are redefined. Recently, asymmetrical parameter-dependent ratios have become popular. Value at Risk (VaR) and Conditional VaR (CVaR) (Martin, Rachev, and Siboulet, 2003) capture downside risk, and Rachev (Biglova, et al., 2004) includes the truncated moments conditioned to tail events as the risk measure.

Few research studies relate portfolio weights directly to state variables. For example, Ait-Sahalia and Brandt (2001) suggest a semi non-parametric method to form a linear index of state variables and select the best variables to directly estimate the optimal portfolio weights. More recently, Laborda and Olmo (2017) provide a linear parametric portfolio policy rule that focuses directly on the dependence of the portfolio weights on the predictor variables. Other related studies are Brennan, Schwartz, and Lagnado (1997), who use a stochastic programming, and Brandt and Clara (2006) and Campbell, Chan, and Viceira (2003), who focus on parametric specifications of the joint dynamics of the state variables and the portfolio returns. The advantage of our approach is that it is a fully non-parametric method that is able to capture nonlinear dependencies between state variables as well as the portfolio weights. This is important, as the dependence of the optimal portfolio choice on the whole return distribution is not linear. We are able to do so by using ANN that has shown excellent performance in pattern recognition in a set of data and is used in various fields. The important feature of ANN, as a machine learning tool, that makes it suitable for this purpose is that it considers patterns including nonlinear and contextual relationships between variables that are often difficult or impossible to detect with linear methods in financial data. In addition to modelling non-linear relations between input and optimal targets and decisions, neural networks can take many input data (e.g., Rojas, 2013 among others). Therefore, the proposed ANN-based method accommodates an arbitrarily large number of state variables in the information set.[4] In few related studies, external features rather than state variables are used in asset allocation and portfolio optimization in neural network settings. For example, Malandri et al. (2018) use public financial sentiment and historical prices to directly prescribe the wealth allocation. They show that with several input variable data, neural networks efficiently learn the impact of public mood on financial time series. Xing et al. (2018) use social media and text mining to translate sentiment information into market view and show that new forecasting techniques improve the portfolio performance. Our study extends this literature by proposing an ANN based approach for state-dependent asset allocation in portfolio management.

The use of ANN as a complex statistical tool improves our portfolio advice, but ANN can be considered as a black box that provides little insight into the influence of state variables on portfolio weights. However, interpretation of statistical models is desired for both academics and practitioners since it helps them understand their portfolio choice in a more intuitive manner. We adopted and applied three methods from computer science to understand better the impact of the state variables on performance ratios as objective functions in portfolio advice.

---

[4] The possibility of incorporating large number of input variables is the feature of ANN as a machine learning tool. Since the proposed method is based on ANN, it allows data to speak and disregard irrelevant state variables in the information set over time. However, in the empirical section of this paper, we use four popular state variables to facilitate comparing the performance of this approach with traditional methods.

The first and second methods, called 'Connection Weights' and 'Permutation Importance', help us to determine the relative impact of each state variable on ratios, but does not guide us as to the direction of the relationship. For this, we use the third method, which is a method of sensitivity analysis called 'Perturb' and is an ideal approach to examine the direction and significance of the impact of an independent variable on the output in a nonlinear model. For brevity, we report and discuss this impact only on Sharpe ratio, as the most common performance ratio used in asset allocation. Our results from the Connection Weights and Permutation Importance methods are qualitatively similar. The results show that, on average, trend (term spread) is the most (least) influential state variable in estimating Sharpe ratio, followed by default spread and dividend yield. The sensitivity analysis using the Perturb method confirms the findings from Connection Weight and Permutation Importance. In addition, it shows that the direction of the impact of trend and dividend yield on Sharpe ratio is different across various sample periods. During the periods that include economic turbulence (1999–2003 and 2004–2008) any changes in trend and dividend yield decrease the Sharpe ratio. The exception is that for trend, over 1999–2003, a small negative change results in a higher Sharpe ratio. During the post-crisis periods (2009–2013 and 2014–2018) both of these state variables have a positive relation with Sharpe ratio. However, the impact is less for the most recent period of 2014–2018 compared to 2009–2013. Interestingly, default spread is the most important state factor during the most recent sample period of 2014–2018.

The contribution of this paper is as follows. Firstly, we suggest a novel approach to conditional asset allocation using ANN, which builds a complex model using state variables to generate optimal portfolio weights directly and without estimation of the return distribution or fitting a model to data. The approach is non-parametric and captures nonlinear dependencies between state variables and portfolio weights. Using ANN makes it possible for the investor to input a range of possible state variables, regardless of their explanatory power on return prediction. ANN distinguishes ineffective inputs and eliminates their additional noise. We show the robustness of our proposed ANN-based method using empirical data and various performance ratios. Secondly, we introduce three analytical methods adopted from computer science that help to interpret the outcome of any machine learning techniques used in finance and economics, such as ANN, by analyzing the influence of input variables on the response variable. This is particularly important, as machine learning tools are criticized in finance and economics fields as being a black box that cannot be used for economic interpretation. Thirdly, we use these three analytical methods and show the relative impact of trend and default spread on Sharpe ratio, as the most important performance ratio used by practitioners in portfolio advice. We also provide evidence that shows that the magnitude and direction of the impact of the state variables in determining Sharpe ratio varies over time. These results suggest that practitioners who use Sharpe ratio should pay attention to these state variables and the sensitivity of their portfolios to the (sometimes mixed) signals that these variables send. The important implication of these findings is that investors should not rely on a single or a specific combination of state variables in their conditional asset allocations since their predictability varies over time. The method we propose in this study is an approach in which we allow data to speak and disregard irrelevant state variables in the information set over time.

The proposed portfolio choice approach is beneficial to managers who aim to efficiently manage a portfolio of different asset classes, and deal with several performance ratios with different return distribution assumptions. This is a challenging task due to the involvement of a large volume of data with different statistical properties, where traditional approaches like linear regression and the generalized method of moments may fail. Ignoring the tail variations of asset returns may result in huge losses; however, having a sufficient knowledge of the relationships between variables results in remarkable gains. In this paper, we address this challenge and propose the use of new statistical approaches and machine learning techniques to construct optimal portfolios based on different market states.

The rest of the paper is organized as follows. In Section 2, we define our portfolio problem as a stochastic optimization problem and discuss our proposed method for portfolio choice using machine learning techniques. Section 3 presents the empirical application of our proposed method, and Section 4 concludes.

## 2. Portfolio Choice Problem

We consider a single-period investor who maximizes the conditional expectation of a performance ratio $\psi(\mathbf{r}_{t+1})$ of next period's portfolio return $\mathbf{r}_{t+1} = \mathbf{x}'_t \mathbf{R}_{t+1}$, where $\mathbf{x}_t$ and $\mathbf{R}_{t+1}$ are the vectors of portfolio weights and returns of the assets, respectively. The expectation is conditional on a vector of state variables $\mathbf{z}_t$. The optimal portfolio weights are obtained by solving the following problem:

$$\max_{\mathbf{x}_t} E[\psi(\mathbf{x}'_t \mathbf{R}_{t+1})|\mathbf{z}_t] \qquad (1)$$

Here, $\psi(\mathbf{r}_{t+1}) = \frac{\phi(\mathbf{r}_{t+1})}{\rho(\mathbf{r}_{t+1})}$ is a performance ratio, and $\phi(\mathbf{r}_{t+1})$ and $\rho(\mathbf{r}_{t+1})$ denote the reward and risk of the portfolio return $\mathbf{r}_{t+1}$, respectively, with two constraints of $\mathbf{x}'_t \mathbf{e} = 1$ and $\mathbf{x}_t \geq \mathbf{0}$. The first constraint describes that no leverage is permitted with $\mathbf{e}$ presenting an unit vector, and the second constraint of non-negativity shows that short selling is not allowed.

When performance ratios are used as the objective function, the return distribution is traditionally assumed to be normal and the parameters or moments of the return distribution are predicted using available data. Therefore, the traditional solution to the investor's problem involves a two-step process (Chen, Tsai, and Lin, 2011; Cornuejols and Tütüncü, 2006). First, the parameters of the returns (mean vector and covariance matrix) are estimated, and second, the estimated parameters are substituted into the performance ratio, and the investor's problem (1) is optimized.

There are various methods to estimate conditional moments of returns. In one common approach, which is unconditional, moments are fitted into a Vector Autoregressive (VAR) model and slopes are used to estimate moments during the out-of-sample period (e.g., Campbell, Huisman, and Koedijk, 2001). Another approach is based on a factor model, in which the time series of the moments is regressed on a set of state variables during the in-sample period. Then, the estimated slopes and state variables are used to estimate the moments during the out-of-sample period (e.g., Chan, Karceski, and Lakonishok, 1999). The main disadvantage of these approaches is that in each of them a particular form for the shape of the returns is assumed. However, the return distribution is rarely known. Also, in the first step of these methods, relationships between variables are assumed to be linear, which results in errors in estimations. In general, the two-step methods lead to a combination of parameter estimation as well as model optimization errors.

In the next section, we propose a method based on ANN, in which a non-linear model directly maps the state variable vector to the optimal portfolio weights without estimation of moments and any distributional assumptions about returns. In this approach, we use ANN over a historical period to develop a model for maximizing the performance ratio of interest that estimates the optimal portfolio weights directly, conditioned on the state of the market. We use this model with the current state variables to estimate the optimal weights for the current period. In the empirical section of the paper, we show the performance of the proposed approach compared to traditional approaches in which the moments of returns are estimated.

## 2.1. ANN for Portfolio Choice Problem

In this section, we briefly review Artificial Neural Network (ANN) as a powerful machine learning tool. Then, we propose a state-dependent model for making decision on portfolio choice using this tool, where the optimal portfolio weights are directly dependent on the market state.

### 2.1.1. ANN

Artificial Neural Network (ANN) is a learning tool for pattern recognition in a set of data. It is, simply, a nonlinear model that has shown excellent performance compared to regression analysis and is used in various fields including Finance (Pyo and Lee, 2018). ANN consists of three layers: input, output and hidden layers which include nodes or neurons, as shown in Figure 1.

[Insert Figure 1 here]

Inputs are multiplied by weights (called network weights or edges) and substituted into functions (called activation functions) in the hidden layer nodes. Activation functions in the hidden layer allow ANN to be highly nonlinear. There are various functions available to be used based on the purpose of the network e.g. tanh, sigmoid, and rectified linear unit (ReLU). Outputs of hidden nodes are multiplied by the weights between hidden and output layers and aggregated into the output nodes using a function (e.g., sigmoid function). The final value of the output layer is used in an objective function that is optimized by changing the weights of the network in various iterations. These weights are optimized by an algorithm such as backpropagation (BP). The number of input and output nodes are determined by the number of inputs in the data sample and the desired number of outputs, respectively. The prediction can improve by increasing the number of hidden nodes, but an excessive number of nodes may lead to an overfitting bias (capturing noise rather than information from the data). The next section describes our proposed approach to use ANN and determine the model that estimates the optimal portfolio weights.

### 2.1.2. Optimal Portfolio Weights and ANN

We use the Lagrangian multiplier method to solve Problem (1). Introducing new variables of $\mu$ and $\lambda$, the Lagrangian function is written as follows:

$$\mathcal{L}_t = E[\psi\left(\mathbf{x'}_t \mathbf{R}_{t+1}\right)|\mathbf{z}_t] + \mu(\mathbf{x'}_t\mathbf{e} - 1) - \boldsymbol{\lambda'}\mathbf{x}_t \qquad (2)$$

The following system of equations, which shows the first order necessary conditions based on partial derivatives of Eq. (2), gives the portfolio weights $(\mathbf{x}_t)$ that solve the optimization problem (1) (Bazaraa, Sherali, and Shetty, 2013):

$$\begin{aligned}
\boldsymbol{\nabla}_{\mathbf{x}_t}\mathcal{L}_t &= 0 \\
\mathbf{x'}_t\mathbf{e} - 1 &= 0 \\
\boldsymbol{\lambda} \odot \mathbf{x}_t &= 0 \\
\mathbf{x}_t &\geq 0 \\
\boldsymbol{\lambda} &\geq 0
\end{aligned} \qquad (3)$$

Here, $\boldsymbol{\nabla}$ is the vector of partial derivatives, and $\odot$ is an element-wise multiplication operator. We know that the function $\psi$ is quasi-concave (as a property of performance ratios) and all the constraints are linear in the portfolio optimization problem. We conclude the following proposition:

**Proposition 1.** *If there exists a unique solution, $(\mathbf{x}_t{}^*, \boldsymbol{\lambda}^*)$, that satisfies system (3), and $\boldsymbol{\nabla}_x\psi(\mathbf{x}_t{}^*) \neq 0$, then $\mathbf{x}_t{}^*$ solves optimization Problem (1).*

This proposition suggests that it is enough to maximize the unconstrained Lagrangian Function (2) rather than optimizing Problem (1). We provide the proof for this proposition in Appendix B. In the portfolio optimization problem, $\mathbf{z}_t$ is a vector of market state variables that captures the market condition. It is well-known in the literature that the selected state variables predict the return moments (see Fama and French, 1988, among others). Ait-Sahalia and Brandt (2001) in a seminal paper show that this relationship also exists between state variables and portfolio choice. In order to incorporate state variables into the Lagrangian function, we relate the portfolio choice $\mathbf{x}_t$ to the state variable vector $\mathbf{z}_t$ by a mapping function $\Lambda(\cdot)$:

$$\mathbf{x}_t = \Lambda(\mathbf{z}_t) \qquad (4)$$

We assume that the function $\Lambda(\cdot)$ is not time variant; however, the time variations of the state variables impose the implied variations in the portfolio choice. Now we suggest a fully nonparametric estimation for $\Lambda(\cdot)$ using ANN, which provides the portfolio choice explicitly as a function of the state variables.

In our proposed ANN, the inputs are state variables $\mathbf{z}_t$, and the sigmoid function[5] is used as the activation function (in the hidden layer). We use the sigmoid function among various kinds of activation functions, since it exhibits the best modeling power. The constraints of Problem (1) impose $\mathbf{x}_t \in [0,1]$, so we also use a sigmoid function in the output nodes that aggregates the outputs from the hidden layer and generates values between 0 and 1 in the output nodes of the network (as the portfolio weights). This allows the non-negativity constraint ($\mathbf{x}_t \geq 0$) to be satisfied automatically, so the Lagrangian Function (2) reduces to:

$$\mathcal{L}_t = E[\psi\left(\mathbf{x'}_t\mathbf{R}_{t+1}\right)|\mathbf{z}_t] + \mu(\mathbf{x'}_t\mathbf{e} - 1) \qquad (5)$$

---

[5] Sigmoid function generates output values between 0 and 1 and is the most popular activation function in multilayer perceptron networks (Alpaydin, 2004).

Here, $\mathbf{z}_t \in \mathbb{R}^M$ where $M$ is the number of state variables, and $\mathbf{R}_{t+1} \in \mathbb{R}^{D_{t+1} \times N}$ where $N$ is the number of assets and $D_{t+1}$ denotes the number of trading days in the investment period.

All historical observations for periods $t = 1 \dots, T$ can be represented as $\{(\mathbf{z}_1, \mathbf{R}_2), \dots, (\mathbf{z}_T, \mathbf{R}_{T+1})\}$. This set of finite historical observations of returns are used to find the reward and risks and compute the ratio and the associated Lagrangian as a function of portfolio weights in each period. The realized asset returns are random variables and their underlying probability is rarely known. Consequently, the Lagrangian is a random function with unknown probability. Therefore, the expectation of the Lagrangian functions over all periods is optimized as the final objective function. In Eq. (5), we substitute portfolio weights $\mathbf{x}_t$ with $\Lambda(\mathbf{z}_t)$, which is the output of the ANN with inputs of state variables $\mathbf{z}_t$ and define the following objective function:

$$\bar{\mathcal{L}} = \frac{1}{T} \sum_{t=1}^{T} \{E[\psi\,(\Lambda(\mathbf{z}_t)'\mathbf{R}_{t+1})|\mathbf{z}_t] + \mu(\Lambda(\mathbf{z}_t)'\mathbf{e} - 1)\} \qquad (6)$$

which is an average of Lagrangian functions $\mathcal{L}_t$'s for all historical sample observations over all periods $t = 1, \dots, T$. The goal of the ANN is to maximize Eq. (6) and find the weights of the network function $\Lambda(\cdot)$. The objective function Eq. (6) is computed using the output of the network which is plugged into both performance ratio $\psi$ and the constraint $\mathbf{x}'\mathbf{e} - 1$. The constraint is multiplied by the Lagrangian multiplier $\mu$ and added to the performance ratio. We incorporate the Lagrangian multiplier $\mu$ into the network as an additional network weight. The network weights are optimized by the backpropagation (BP) algorithm, which is the most common method, used in almost 80% of all applications (Kaastra and Boyd, 1996). In the optimization algorithm, we use the gradient descent approach, which follows the positive direction of the gradients to optimize Function (6)[6]. Using this algorithm, we optimize the objective function iteratively by generating a sequence of network weights based on random gradients until there is no improvement in the function[7]. If the iteration process converges, we are sure that the budget constraint is satisfied[8] (Rojas, 2013); i.e., $\mathbf{x}'_t\mathbf{e} = 1$. Once the network is trained[9] (after the last iteration), we have the model $\hat{\Lambda}$ that enables us to obtain portfolio weights for a new period, directly, from the network's inputs (i.e., state variables). The proposed ANN, as a model to find the optimal weights given market states, is shown in Figure 2.

---

[6] The aim of the stochastic gradient descent method (Hannah, Powell, and Blei, 2010) is to optimize a function iteratively by generating a sequence of solutions based on the estimated gradients calculated from a randomly selected subset of the data. The algorithm stops when there is no improvement in objective function, or some other criteria are met. The learning rate is considered as a step decaying at each iteration of the algorithm computed as $\gamma_i = \gamma_0/1 + c_i$, where $\gamma_i$ is the rate of $i^{th}$ iteration, $c_i$ is the number of iteration, and $\gamma_0$ is a constant. We do cross validation to choose the best parameters on the training set.

[7] We control the number of nodes and other associated parameters by implementing cross validation and measuring the objective function value over training process to ensure we avoid the overfitting problem. Technical details are available upon request.

[8] To maximize function (6) its gradients have to be set to zero. Here, we do this numerically and we need to follow the positive gradient direction to find a local maximum point. Since $\frac{\partial \bar{\mathcal{L}}}{\partial \mu} = \mathbf{x}'_t\mathbf{e} - 1$ is one of the gradients and it is zero in optimality, so the algorithm does not end as long as $\mathbf{x}'_t\mathbf{e} - 1 \neq 0$.

[9] Training refers to the process of modeling the observation data by machine learning tools like ANN to infer a function. The observation data used in the process and the final estimated model are known as 'training data' and a 'trained model', respectively.

[Insert Figure 2 here]

It should be noted that according to the theory of learning and the principle of empirical risk minimization, the risk function in the learning process can take any form of loss, risk, or cost function including simple forms such as root mean square error (RMSE) (Vapnik, 1992). We do not use loss functions such as RMSE because we do not use the optimal portfolio weights as targets in the learning process, instead we maximize the performance ratios numerically and through the learning process. In this setting, we incorporate the budget constraint into the objective function, and the new unconstrained Lagrangian function is considered for the learning process. To optimize this function numerically, we follow the gradient direction to find the optimum using partial derivatives and updating the network weights iteratively that provide the optimal portfolio weights as the network's output (see, Rojas, 2013 and Oroojlooyjadid et al., 2020). This method ensures that the learning proceeds in the direction of minimizing the distance between the final portfolio weights and the unobservable optimal weights.[10]

Algorithm (1) provides the training procedure of the proposed ANN based on the iterations required in the optimization algorithm.

| Algorithm 1: ANN training for the training period |
| --- |

1 **Start**
2   with the ratio $\psi$ **do**
3     $W, \mu \leftarrow$ random initial weights for the Network $\Lambda$
4     **while** stopping criterion is not met **do**

5         **for** $t \leftarrow 1$ **to** $T$ **do**
6             observing the vector of state variables $\mathbf{z}_t$
7             $\mathbf{x}_t \leftarrow \Lambda(\mathbf{z}_t)$, obtaining portfolio weights from the network
8             $\mathbf{r}_{t+1} \leftarrow \mathbf{x}'_t \mathbf{R}_{t+1}$, computing portfolio returns using realized returns $\mathbf{R}_{t+1}$
9             $\mathcal{L}_t$, computing the Lagrangian function
10         **end**

11         calculating the objective function $\bar{\mathcal{L}}$ (Eq. (6)) using all $\mathcal{L}_t$'s
12         finding partial derivatives $\frac{\partial \bar{\mathcal{L}}}{\partial W}$ and $\frac{\partial \bar{\mathcal{L}}}{\partial \mu}$
13         $W \leftarrow W + \frac{\partial \bar{\mathcal{L}}}{\partial W}$ and $\mu \leftarrow \mu + \frac{\partial \bar{\mathcal{L}}}{\partial \mu}$, updating network weights
14     **end**

15     **return** trained network $\hat{\Lambda}$
16 **Stop**

---

[10] In other words, we use ANN as an optimization tool which has been discussed in the computer science literature and used in few fields such as inventory, logistics and smart grids (e.g., Oroojlooyjadid et al., 2020; and Villarrubia et al., 2018).

So far, we have shown how to train the ANN and find the weights of assets in a portfolio that maximize a specific performance ratio. We use the trained ANN and the state variables, for which the information is available at the beginning of the period, to estimate optimal portfolio weights in the new period $T+1$ ($\mathbf{x}_{T+1}$). More formally:

$$\mathbf{x}_{T+1} = \hat{\Lambda}(\mathbf{z}_{T+1}) \qquad (7)$$

In this section, we proposed an ANN-based method for state-dependent asset allocation which estimates optimal portfolio weights directly by using state variables without assuming any distribution of returns and other data, without fitting a model with a fixed number of predicting variables to data and without estimating any parameters. Since the method is based on machine learning, it accommodates any number of state variables in the information set. This is possible because the network of the machine learning disregards irrelevant state variables. Next, we test the proposed method empirically and compare its performance to that of traditional methods using various performance ratios.

## 3. Empirical Application

### 3.1. Data and Variable Construction

We construct portfolios using the Standard and Poor (S&P) 500 Index and the US benchmark thirty-year government bond index. Daily returns on these indices are sourced from Refinitiv Thomson Reuters and cover the period from the beginning of January 1986 to the end of December 2018.

There are various economic variables that have been shown to partly capture the state of the market and predict the moments of returns[11]. We follow Ait-Sahalia and Brandt (2001) and collect monthly data on four popular state variables: the default spread, the dividend yield (of the S&P 500 Index), the term spread, and an S&P 500 Index trend (or momentum). They show that these variables capture time variations of at least first and second moments of bond and stock returns and are good candidates as predictors.

The default spread is the monthly difference between Moody's BAA- and AAA-rated corporate bond yields. The term spread is the monthly difference between the rates on 10-year and 1-year Treasury Constant Maturity Rates. The data to construct term spread and default spread are from the Federal Reserve Bank of St. Louis during our sample period. The dividend yield is the log dividend-to-price ratio of the S&P 500 Index, where the dividend-to-price ratio is the sum of dividends paid on the S&P 500 Index over the past 12 months divided by the current level of the Index. We download monthly aggregate dividend-to-price data over our sample period from Robert Shiller's website, http://aida.econ.yale.edu/_shiller/. The trend is the log of the ratio of the current S&P 500 Index level and the average Index level over the previous 12 months.

---

[11] Some of the variables on mean predictability include term spread (Campbell, 1987; Fama and French, 1988, 1989; Ferson and Harvey, 1991), default spread (Fama and French, 1988, 1989; Keim and Stambaugh, 1986) and Treasury bill yield (Fama and Schwert, 1977; Ferson and Harvey, 1991). The variables for predicting variances include lagged squared return and/or lagged variance (Bollerslev, 1986; Engle, 1982; French, Schwert, and Stambaugh, 1987; Harvey, 2001; Schwert, 1989; Whitelaw, 1994), default spread (Harvey, 2001; Whitelaw, 1994), dividend yield (Harvey, 2001) and debt-to-equity ratio (Schwert, 1989). Finally, the predictability of convarinces is attributed to variables such as lagged covariances, lagged cross-products of returns (Bollerslev, Engle, and Wooldridge, 1988), term spread (Campbell, 1987; Harvey, 2001), and default spread and dividend yield (Harvey, 2001).

The sample includes 8,609 daily observations of the S&P 500 and bond indices. The monthly data of the state variables contains 396 observations. Figure 3 and Table 1 show the data over the 33-year sample period.

[Insert Figure 3 here]

[Insert Table 1 here]

Table 1 shows that the S&P 500 Index has a higher average and standard deviation than those of the bond index, as expected. Both indices have a small, but negative skewness. Kurtosis for the S&P 500 Index suggests that it is very fat-tailed compared to the normal distribution, while the bond index has smaller kurtosis in excess of a value of 3. We also ran the Jarque–Bera (JB) normality test (results are not reported) for the returns on S&P 500 and the bond indices and reject the null hypothesis that returns are normally distributed.

### 3.2. ANN-based approach versus Traditional Optimal Portfolio Choice

As the objective function in a portfolio choice, various performance ratios are discussed in the literature. In this study, we choose six popular performance measures[12] to confirm the outperformance of the ANN-based approach compared to traditional methods. However, our machine learning procedure is not restricted to the type of performance measure. We use Sharpe, Mean Absolute Deviation (MAD)(Konno and Yamazaki, 1991), MiniMax (Young, 1998), Gini (Shalit and Yitzhaki, 1984), CVaR (Martin, Rachev, and Siboulet, 2003), and Rachev (Biglova et al., 2004) ratios.[13] These performance measures are widely used in the literature (e.g., Biglova et al., 2004; Farinelli et al., 2008; and Ortobelli, et al., 2018) and include both symmetric and asymmetric measures. Appendix A provides a summary of these measures and how they are constructed.

We use rolling 156-month (13-year) observations from January 1986 as a *training* set (or in-sample) and test the trained model on the following 60 months as a *test* set (or out-of-sample).[14] We roll over the training and out-of-sample sets every 60 months. Figure 4 shows the rolling windows of training and out-of-sample sets.

---

[12] We use two terms of 'performance measure' and 'performance ratio' interchangeably.

[13] We consider CVaR with parameter $\alpha = 0.5$, and for Rachev we use the parameter $\alpha = 0.5$ and $\beta = 0.99$. However, our proposed procedure is not sensitive to these settings.

[14] In empirical applications of machine learning tools, it is common to divide the whole sample into two parts, in-sample and out of sample, which usually contain 70% and 30% of the observations, respectively.

[Insert Figure 4 here]

This approach provides four 5-year consecutive out-of-sample periods starting in January 1999 until the end of our sample in December 2018[15]. We did not break our sample simply into two subsamples of training and out-of-sample, as the rolling windows allow us to include the most recent information in the model. We did not roll the windows more frequently than every five years, as a shorter frequency is unlikely to provide more information to the trained model, given we use monthly state variables.

We apply our designed ANN to construct monthly portfolios given the state of the market. As the first step, we use state variables at the beginning of each month and daily returns of the stock and bond indices during the month, over the training sets, and train an ANN using Algorithm (1). These trained ANNs are the models (for each performance ratio) that take the state variables at the beginning of the (new) month[16] as the input and give an estimation for the optimal portfolio weights for that month as the output. These are the weights that maximize the associated ratio for a given month, conditioned on the market state. Algorithm (2) describes these steps for testing in each rolling window.

---

Algorithm 2: ANN for testing during out-of-sample period

---

1 **Start**
2  with each ratio $\psi$ **do**
3   load the trained network $\hat{\Lambda}$ from Algorithm (1)

4  **for** $t \leftarrow T+1$ **to** end of out-of-sample period **do**
5     observing the vector of state variables $\mathbf{z}_t$
6     $\mathbf{x}_t \leftarrow \hat{\Lambda}(\mathbf{z}_t)$, obtaining portfolio weights from the network
7     $\mathbf{r}_{t+1} \leftarrow \mathbf{x}'_t \mathbf{R}_{t+1}$, computing portfolio returns using realized returns $\mathbf{R}_{t+1}$
8     $\psi(\mathbf{r}_{t+1})$, calculation of the ratio with portfolio returns
9  **end**

10  **return** out-of-sample series of ratio values
11 **Stop**

---

It is important to investigate whether using ANN in the maximization of the ratios outperforms the traditional methods. We employ three traditional methods, as benchmarks. In the first method, moments of returns are fitted into a Vector Autoregressive (VAR) model over the in-sample periods, and then slopes are used to estimate moments and maximize performance ratios during the out-of-sample period[17]. This approach is unconditional since state variables are not used in the estimation of moments. In the second method, which is conditional, a factor model is used to regress time series of the moments on a set of state variables during the in-sample period. Then, the estimated slopes and state variables are used to estimate the moments and maximize performance ratios during the out-of-sample period.[18] In the third method, a parametric method proposed by Brandt et al. (2009) is used to estimate optimal weights directly from the state variables. In this benchmark, an expected utility function is maximized during in-sample period to obtain the coefficients in a linear model that relates state variables to portfolio weights. Then, the estimated coefficients and state variables are used to estimate the portfolio choice during the out-of-sample period. In this approach the expected utility, as the objective function, would implicitly take into account the different moments of the return distribution and the risk profile of the investor. The details and specifics of these three benchmark methods that we use in this study are as follows.

---

[15] Out-of-sample periods are 1999–2003, 2004–2008, 2009–2013 and 2014–2018. For each out-of-sample period, ANN is trained over the previous 156 months of observations.

[16] The information for state variables is available at the end of the previous month.

[17] For example, see Chen, Tsai, and Lin (2011), Campbell, Huisman, and Koedijk (2001), Kroll, Levy, and Markowitz (1984) and Bekaert and Hodrick (1992) among others that use the same approach.

*Benchmark method (1):* first, we calculate monthly moments (mean vector and covariance matrix) of assets using daily returns over the in-sample periods, and then, these monthly moments are fitted into an autoregressive process with one lag (AR (1)) as a VAR model[19] over the in-sample periods. During the out-of-sample periods, the estimated slopes and the realized moments (as input) are used to estimate monthly moments. We use the monthly estimated moments to simulate daily returns[20] of the assets and consequently maximize the performance ratios[21] to determine the optimal weights per month and during the out-of-sample period.

*Benchmark method (2):* as a first step, we calculate monthly moments (mean vector and covariance matrix) of assets using daily returns over the in-sample periods. Then, we use a factor model that includes the four state variables that we have used in this paper (i.e., the default spread, the dividend yield, the term spread, and the S&P 500 Index trend) and regress monthly moments on the set of state variables during the in-sample period. During the out-of-sample periods, the estimated slopes and state variables (as input) are used to estimate monthly moments. We use the monthly estimated moments to simulate daily returns of the assets and consequently maximize the performance ratios to determine the optimal weights per month and during the out-of-sample period.

*Benchmark method (3):* following Brandt et al. (2009), we consider the Constant Relative Risk Aversion (CRRA) utility function with a prespecified risk parameter, $\gamma = 5$.[22] We then optimize the expected utility function over in-sample periods to obtain the optimal coefficients in a linear model that makes the portfolio weights dependent on the state variables. In out-of-sample periods, we use the estimated linear model and the new state variables at the beginning of each period to find the optimal portfolio weights per month and during the out-of-sample period.

---

[18] See Chan, Karceski, and Lakonishok (1999) for the approach among others.

[19] Our results are not sensitive to the choice of lags.

[20] The returns of assets are assumed to have a multi-variate normal distribution. We follow the simulation method suggested by Kotz et. al (2000) to simulate daily returns.

[21] In order to maximize Sharpe, Gini and Rachev, we follow optimization approaches suggested by Cornuejols and Tütüncü (2006). We use Konno and Yamazaki's (1991) method to maximize MAD, and finally we follow Rockafellar and Uryasev's (2002) approach to optimize CVaR and MiniMax. In general, in all these approaches, the idea is to iteratively explore the portfolios at different return levels on the efficient frontier and locate the one with maximum ratio.

[22] We also examine the risk parameter $\gamma = 100$ in benchmark (3) for investors who are extremely sensitive to losses (Brandt et al. 2009). The performance ratios with this risk parameter are significantly lower compared to the risk parameter $\gamma = 5$.

To compare these benchmark methods with our proposed method, we use the monthly estimated optimal portfolio weights from each benchmark and realized asset returns to compute portfolio returns and performance ratios during the out-of-sample periods. The in-sample periods used in benchmark methods (1) and (2) to calculate the moments of returns; and in benchmark method (3) to maximize the utility function and estimate the linear model are identical to those used to train the ANNs. For benchmarks (1) and (2), we maximize the performance ratios during similar out-of-sample periods which is a total of 240 months. For the benchmark (3), we use the estimated linear model and state variables to obtain the optimal portfolio weights during similar out-of-sample periods which is a total of 240 months. We also use trained ANNs over the same 240-month of out-of-sample periods.

Table 2 reports the descriptive statistics of the obtained values for each performance ratio using these four methods across the whole out-of-sample period of 1999–2018. It also shows the significance level of the mean difference between the values of the ratios maximized using ANN-based method and the benchmark method that generates the higher value (i.e., the best benchmark). The means of all ratios are greater when we use ANN to find the optimal weights compared to other methods. For example, using ANN-based approach, the average Sharpe ratio (CVaR ratio) is 0.09 (0.34) on monthly basis, whereas it is 0.08 (0.22) based on the benchmark method (3), 0.07 (0.20) based on the benchmark method (2), and 0.06 (0.17) based on the benchmark method (1). The biggest improvement is for CVaR and MAD ratio, which increases from 0.22 and 0.10 under the benchmark method (3) to 0.34 and 0.15, respectively, using ANN-based approach. This indicates about 50% improvement in their performance. For Rachev, there is an increase from 3.44 under the benchmark method (3) to 4.45 using ANN-based approach.[23]

[Insert Table 2 here]

Additionally, of the three benchmarks, the greater ratio value is obtained by the benchmark method (3), which is an integrated and parametric linear model in which state variables are directly connected to portfolio weights. The second-best method is benchmark method (2), a conditional approach, compared to the unconditional benchmark method (1) where the portfolio problem is solved in two steps. These results show the importance of considering the state of the market in the asset allocation problem as well as integration of estimation and optimization steps. However, the factor model used in the benchmark method (2) and the integrated version in benchmark method (3) assume that the relationship between return moments and market state variables is linear. Our results show that ANN provides a better outcome as it captures the nonlinear relations between these variables and the portfolio weights using an integrated approach. In other words, improvement for all ratios using the ANN-based approach compared to the benchmarks is not only due to considering state of the market in portfolio choice or connecting state variables directly to the estimated weights. Table 2 also reports other statistics of each performance ratio. Standard deviations of values of ratios obtained by ANN are relatively close to benchmarks, while the negative skewness disappears for Sharpe, CVaR, and Rachev. Skewness and kurtosis increase using ANN compared to benchmark methods (1) and (2) for all ratios except for Rachev, for which these values are close. The skewness and kurtosis obtained by benchmark method (3) are larger than ANN method in all cases except the skewness in CVaR and Rachev, and kurtosis in Rachev.

---

[23] In addition to these three traditional benchmarks, we also consider "static" portfolios with static weights for bond and stock portfolios which are common among practitioners. We construct three static portfolios as Benchmark-static$_{x\text{-}y}$, where x and y denote the weights (in percent) of the S&P 500 index and the bond index in the portfolios, respectively. We use the static weights and asset returns to compute portfolio returns and performance ratios during the out-of-sample periods. The mean values of different performance ratios for these static portfolios are as follows. Benchmark-static$_{20\text{-}80}$: Sharpe=0.05, MAD=0.06, MiniMax=0.04, Gini=0.09, CVaR=0.18, Rachev=3.31; Benchmark-static$_{60\text{-}40}$: Sharpe=0.07, MAD=0.09, MiniMax=0.05, Gini=0.12, CVaR=0.25, Rachev=3.56; and Benchmark-static$_{80\text{-}20}$: Sharpe=0.07, MAD=0.09, MiniMax=0.06, Gini=0.12, CVaR=0.17, Rachev=3.53. Comparing these results with those in Table 2 indicates that the means of all ratios are greater when we use ANN to find the optimal weights compared to the static approach. We thank the associate editor for this valuable suggestion.

To demonstrate the performance of the ANN-based method compared to benchmark methods over time, for each year we compute the monthly average of each ratio using different methods. We show the time series of means in Figure 5. The plots show that in almost all years and across all ratios the ANN-based method provides higher values for performance ratios.

[Insert Figure 5 here]

We further present the performance of the ANN-based approach during subsamples of the whole out-of-sample period: i.e., 1999–2003, 2004–2008, 2009–2013 and 2014–2018 in Table 3. These periods are associated with important economic and financial turbulence. The 1999–2003 period includes the Dotcom bubble, 2004–2008 started with a bull market and ended with the Global Financial Crisis (GFC), and 2009–2013 is associated with the post-GFC period. Finally, 2014–2018 includes the recent growing period. The results from subsamples confirm our findings from the whole out-of-sample period. The average value of each ratio is higher when we use ANN compared to three benchmark methods. In almost all subsamples and for all ratios, the difference between mean values of the ratios using ANN and the best benchmark is positive and significant. This suggests that the proposed approach is robust in outperforming other traditional methods in different episodes of the market.

[Insert Table 3 here]

To get another insight into comparison of the proposed approach and other benchmarks, we analyze the portfolio weights over the whole out-of-sample period for the ANN-based approach as well as all other benchmark methods.[24] Figure 6 shows the average of the monthly estimated optimal weights of the S&P 500 index over each year from 1999 to 2018. For brevity we present the analysis for benchmarks (1), (2) and ANN-based methods using only Sharpe ratio.[25] The weight for the Bond index is 1- the weight of the S&P 500 index. The results show that the weights on the equity index using ANN-based method and its closest benchmark i.e., benchmark (3) which is based on Brandt et al. (2009), follow the same pattern. The weights on equity index decrease during the Dotcom bubble and the GFC and increase afterwards as expected. However, during the post-GFC period, the variations in portfolio weights are larger when we use benchmark (3) compared to the ANN-based method, indicating that the proposed ANN-based approach can capture the information set better than the benchmark model. The patterns of the optimal weights for benchmarks (1) and (2) are almost similar over time. In both methods, the weights on equity decrease during the Dotcom bubble and the GFC and increase afterwards, however, the rates of these changes are different in each method. Also, the weights on equity index, on average, are less than 50% which lead to lower performance of these benchmarks compared to other methods as reported in Table 2.

---

[24] We thank an anonymous referee for this suggestion.
[25] The results for other ratios are qualitatively similar and available upon request. Also, note that in benchmark (3), the objective function is a utility function.

[Insert Figure 6 here]

In this study, we have used four state variables as the determinants of the optimal portfolio weights. It is interesting to examine the relative impact of each state variable on performance ratios. Since ANN is a nonlinear approach, a linear regression of time series values of the computed ratios on the lagged state variables does not reveal the actual relations. In the next section, we use analytical methods from computer science to understand better the impact of the state variables on the performance ratios. For brevity, we report and discuss this impact only on the Sharpe ratio, as the most common performance ratio used by practitioners and academics in finance.[26]

### 3.3. Sharpe Ratio and State Variables

The use of new statistical methods such as ANN improves prediction and estimation for many different problems. This advantage is due to the complexity of these methods and their ability in nonlinear modeling (Olden and Jackson, 2002). However, neural networks such as ANN are considered as a 'Black Box' with excessive connection weights, which provide little insight into the influence of input variables on the response variable. Nevertheless, interpretation of statistical models is desired in many fields, such as finance, where practitioners seek knowledge of relationships, leading to behavior or pattern detection.

Several methods have been introduced in computer science that identify the importance of explanatory variables and their relative contributions to the modeling process in neural networks, yet to be used in Finance applications. We use two methods to find the relative importance (RI) of the state variables in obtained values of the Sharpe ratio. The first method is called 'Connection Weights' which has been shown to be the most successful to determine the importance of independent variables in a network (Olden and Jackson 2002) and calculated using the estimated networks weights.[27] The second method is the 'Permutation Importance' method which is common in the random forest literature (Breiman, 2001) and helps us to measure the relative importance of state variables by analysing the impact of the permutation of the values of a state variable on the portfolio using out-of-sample data.[28] In our sensitivity analysis, each method can be used as the robustness for another approach. Using both methods, we can provide better insight with respect to the relative importance of a state variable in the value of the Sharpe ratio.

---

[26] Results for other performance ratios are available upon request.
[27] The alternatives are Partial derivatives (PaD) (Dimopoulos et al., 1995, 1999), Garson's algorithm (Garson, 1991), Perturb method (Yao et al., 1998; Scardi and Harding, 1999) and Profile method (Lek et al., 1996a, b). Olden, Joy, and Death (2004) shows Connection Weights performs better than other methods.
[28] We thank an anonymous referee for this suggestion.

In 'Connection Weights' approach, the RI for an input variable is computed as the sum of the products of *final* network weights coming in and out of hidden nodes of the network. For each state variable, we calculate the RI measure as:

$$RI_i = \sum_{h=1}^{H} w_{ih}.w_{ho} \qquad (8)$$

Here, $w_{ih}$ and $w_{ho}$ indicate the final network weighs between state variable $i$ and hidden node $h$, and the network weight between hidden node $h$ and the output node $o$ (portfolio weights used to estimate Sharpe ratio)[29], respectively. We use the estimated network weights of the ANNs for each subsample and compute the RI measure for each state variable.

In the 'Permutation Importance' method, the trained networks and the original out-of-sample input variables are used to compute a reference score based on the outputs. Then, an input variable is shuffled for several repetitions and the associated scores are generated. The RI for the input variable is calculated as the difference between the average of the scores from the repetitions and the refence score. We use the trained networks for each subsample and the original out-of-sample state variables to compute the reference score that is the value of Sharpe ratio, $s$. We then shuffle a state variable $i$ to generate a new version of the state variable for $K = 100$ repetitions. The score (Sharpe value) of the network for each repetition based on all state variable including the permuted state variable $i$ is computed as, $s_{k,i}$. The Importance of state variable $i$ is calculated as follows:

$$RI_i = s - \frac{1}{K} \sum_{k=1}^{K} s_{k,i} \qquad (9)$$

In both methods, the larger (smaller) value of RI indicates the higher (lower) importance of the associated state variable in determining portfolio weights and consequently estimated Sharpe ratio.

[Insert Table 4 here]

Table 4 reports the values for RIs for 'Connection weights' and 'Permutation Importance' methods in panels A and B, respectively. The results from both methods are almost consistent over all subsamples as well as the whole period and give us similar relative importance rankings for the state variables. In our analysis, if the RI values of state variables are equal in one method, we consider the other approach to identify the more influential state variable. The results in Table 4 show that trend is the most important state variable for the Sharpe ratio across all subsamples except for the sample period of 2014–2018 during which default spread is the most important one followed by trend. Over the sample periods of 1999–2003 and 2004-2008, default spread is the second most important variable, whereas during sample period of 2009–2013, dividend yield is the second most important variable. In summary, Table 4 shows that trend (term spread) is the most (least) influential state variables in estimating the Sharpe ratio, followed by default spread and dividend yield.

[29] Since we have only two asset classes in our empirical section, our network has one output node which is the portfolio weight (x) for one asset class (the portfolio weight for the other asset class is 1-x).

The relative importance approach does not give us any indication of the direction and magnitude of the impact of variables on the Sharpe ratio. To do so, we conduct a sensitivity analysis called the 'Perturb' method, which is an ideal approach to examine the direction and significance of the impact of an independent variable on the output in a complex, nonlinear model with various independent variables.

The sensitivity analysis is a visual presentation showing the impact of small changes in each parameter of a model on the value of the function (Sharpe ratio) in question, all on the same graph. In our analysis, values of a state variable, as the input of the ANN, are varied across a range of values and the effect on Sharpe ratio is observed in each scenario. The values of other state variables are the actual data. Then we display the range of changes in Sharpe ratio against the changes in each state variable. In our analysis, we change the values of a state variable by $\pm 3\sigma$, $\pm 2\sigma$, and $\pm \sigma$, where $\sigma$ is its standard deviation, and calculate the Sharpe ratio using the ANN over the out-of-sample periods. We plot the percent change in the estimated Sharpe ratio against these variations for each state variable and subsample, as shown in Figure 7.

The Y axis shows the expected range of Sharpe ratio, if the value of either trend, default spread, term spread or dividend yield is shifted somewhere between a minimum and maximum of the fixed range we have defined based on their deviations from their means (standard deviations). The plots on this figure show how Sharpe ratio increases or decreases as the value of one state variable is changed and all others held to their actual values. The plots reveal the nonlinear relationships and the relative sensitivity of the Sharpe ratio to changes in each state variable.

[Insert Figure 7 here]

Figure 7(a) shows the plot for the 1999–2003 period which includes the Dot-com crash that lasted from March 2000 to October 2001. The largest Sharpe ratio range is for trend, followed by dividend yield, and positive changes in all state variables have a negative impact on Sharpe ratio. Negative changes of state variables have a different impact on the Sharpe ratio. The negative changes of default spread and term spread (dividend yield) decrease (increase) the Sharpe ratio, while small (large) negative changes in trend increase (decrease) the Sharpe ratio.

The plot for trend and dividend yield also reveals that a positive change in trend has a more pronounced negative effect on the Sharpe ratio than a positive change in dividend yield. The direction of the effect is opposite for the extreme negative changes in trend and dividend yield. The extreme negative changes in trend (dividend yield) decrease (increase) the Sharpe ratio.

The plots from the period 2004–2008 (Figure 7(b)), which includes the GFC, provide similar but stronger results. The Sharpe ratio is more sensitive to trend, followed by dividend yield and default spread. Furthermore, any positive or negative changes in any state variable have a negative impact on the Sharpe ratio. This negative effect is equally pronounced for both positive and negative variation in state variables. During this period default spread and dividend yield have very similar effect on the Sharpe ratio.

Figure 7(c) shows the sensitivity plots for post-GFC period (2009–2013). Consistent with results in panels (a) and (b), trend and dividend yield have the largest impact on the Sharpe ratio. However, there is a positive relation between these state variables and the Sharpe ratio, and their impact is large. For example, one standard deviation variation in these variables increases the Sharpe ratio by about 100 percent. Term spread has a negative relation with Sharpe across the full range of variations in its value, while default spread has a positive relation with Sharpe for the extreme changes in its value.

The sensitivity plots for the period 2014–2018 in Figure 7(d) are similar to those for the post-GFC period in panel (c) where the relations between trend and dividend yield and Sharpe ratio are both positive. However, their impact on Sharpe ratio is less than for the post-GFC period, particularly for dividend yield, for which the Sharpe ratio varies less than 10 percent across its full range of variation. During this period, the largest Sharpe ratio range is for default spread, followed by trend. Positive variation in default spread brings down the Sharpe ratio by close to 50 percent, while the extreme negative changes in default spread have less impact on the Sharpe ratio. Term spread has a negative relation with Sharpe ratio and its impact on Sharpe ratio is less, compared to that of the post-GFC period.

In general, sensitivity plots in Figure 7 suggest that trend (term spread) is the most (least) important state variable for the Sharpe ratio compared to other state variables. The direction of the impact of term spread on Sharpe ratio is different across sample periods. During the periods that include economic turbulence any changes in term spread values decrease the Sharpe ratio, while during the post-crisis periods term spread has a negative relation with this ratio. Dividend yield and trend have a similar impact on the Sharpe ratio over these crises, and their relations with Sharpe are positive during the post-GFC and recent periods (2014–2018); however, the impact of dividend yield is not as strong as trend for the most recent sample period (2014–2018). The changing range for the Sharpe ratio is similar for both default spread and dividend yield for the subsample periods that include the economic crises. However, during the post-GFC period, dividend yield is a more important determinant of Sharpe ratio compared to default spread. Nevertheless, during the most recent period of 2014–2018, default spread is the most important state factor.

Our results from the sensitivity analysis are consistent with the findings in Table 4, which show trend as the most important determinant of Sharpe ratio, followed by default spread and dividend yield. They also provide further support for the finding that default spread is the most important variable for the recent period of 2014–2018.

In summary, in this section we find that the trend (or momentum) of the market and default spread are the most important determinant of the Sharpe ratio, as an objective function, in portfolio choice, but that the direction of their impact on the Sharpe ratio varies over time. These results suggest that managers who use the Sharpe ratio as their objective function should monitor these state variables and the sensitivity of their portfolios to the variation in these variables. The broader insight from these findings is that investors should not rely on a single or a specific combination of state variables in their portfolio advice since their effect varies over time. This also highlights the advantage of the proposed method, which is not restricted to the type or number of state variables in the information set.

### 3.4. Time varying performance of ratios

The question of which performance ratio is the best for portfolio management has long been debated. We employ the optimal portfolio choice for each performance ratio estimated by ANN over the out-of-sample period to shed some light on this issue. In particular, we investigate if there is a single performance ratio that outperforms other ratios for all periods by looking at the final wealth of associated optimal portfolios that we obtained using our ANN-based approach. The aim is not to identify which ratio amongst our six ratios is the best; rather to examine whether there is evidence of a single best ratio across all periods.

For each ratio during the out-of-sample period (240 months), we use the optimal weights based on the ANN-based method and compute monthly returns. Table 5 reports the average monthly returns for each ratio in panel (a) and the number of months that each ratio generated the highest return (final wealth) compared to other ratios in panel (b). The results from panel (a) show that Rachev and MiniMax ratios generate higher monthly returns compared to other ratios, while the Sharpe ratio earns the lowest return. However, Rachev and MiniMax are not the absolute outperformers. Panel (b) shows that MiniMax and Rachev are the best ratios for 67 and 42 months out of 240 months, respectively, and other ratios perform better in other months. Interestingly, Sharpe is the best ratio for only 23 months and is ranked last of the six ratios. CVaR is only slightly more frequently successful than Sharpe with 27 months as the best ratio, but it ranks third for performance.

[Insert Table 5 here]

In Figure 8, we show which ratio outperforms the others in each month. The success of MiniMax, which is the best ratio for 67 months, are concentrated mostly during the periods of 1999–2008 and 2014–2018. This ratio is the best ratio during and after the Global Financial Crisis (GFC). On the other hand, CVaR and MAD perform the best during the GFC, while Rachev and Gini outperform other ratios in the post-GFC period.

These findings show that frequency and pattern of success for a performance ratio are not good indicators of the final outcome based on that ratio. In other words, their performance is time-varying.

To confirm this, we investigate the relationship of the frequency of success of a ratio with its returns. We rank the ratios with respect to frequency as well as the monthly returns and report the Pearson's correlation between the ranks in Table 5. The correlation is insignificant at conventional levels, suggesting that there is no relationship between these rankings. These results are different from the findings of Farinelli et al. (2008), who show that the leading ratios in terms of producing returns are the most frequently successful ratios. Our results in Table 5 suggest that the frequency of success of ratios is not a credible rule to guide the managers and they should not focus on a single performance ratio in their management process.

[Insert Figure 8 here]

## 4. Conclusions

In this paper, we contribute to the literature on state-dependent asset allocation by proposing a novel procedure using ANN to determine conditional portfolio choice. Instead of first modeling various features of the conditional return distribution and subsequently estimating the portfolio choice, which is a common approach, we focus directly on the dependence of the portfolio weights on the state variables. The empirical results for a portfolio of bond and stock indices show that the suggested method of portfolio choice results in a better portfolio composition with a greater value for various performance ratios than that of traditional approaches.

The presented results have important implications for portfolio managers. While the dynamic adjustment of the allocations results in more optimal portfolios compared to unconditional methods, the outcome can improve significantly when the non-linearity between state variables and portfolio weights is considered. The proposed method is a new robust approach, based on machine learning that improves the risk-adjusted return by taking into account all the dependencies between variables, and relates portfolio weights directly to the state variables. This is important as financial data is inherently noisy, and the relationships between the state variables and returns can be variable, nonlinear, and/or contextual.

Furthermore, we employ novel methods that suit complex nonlinear statistical models used in neural networks and analyze the relative impact of state variables on Sharpe ratio, a common ratio used in asset allocation. We find that trend and dividend yield are the most important determinants of the Sharpe ratio. However, the direction and magnitude of their impact change over time. The findings suggest that it is not reliable to use a single or a specific combination of state variables in conditional asset allocations, as their predicting power may change while the dynamics of the market changes over time.

In this study, we considered a single-period (myopic) investor who maximizes her next-period objective. As a future research, our procedure can be extended to a multi-period asset allocation for an investor who has preferences over terminal wealth using other machine learning tools that are able to cover time stages in a long-term investment horizon.

# Appendix

## Appendix A: Performance Ratios

**Table A.1**: Performance ratios with rewards and risk measures and their parameters.

| Performance Ratio | Parameters |
|---|---|
| Sharpe (Sharpe, 1966): $E[r_p]/\left(E[(r_p - E[r_p])^2]^{\frac{1}{2}}\right)$ | - |
| MAD (Konno and Yamazaki, 1991): $E[r_p]/E[\lvert r_p - E[r_p]\rvert]$ | - |
| MiniMax (Young, 1998): $E[r_p]/\left(\inf\limits_{t\in\{1,D\}}(r_{p_t})\right)$ | - |
| Gini (Shalit and Yitzhaki, 1984): $E[r_p]/\left(\frac{1}{2}E\left[\lvert r_{p_t} - r_{p_{t'}}\rvert\right]\right)$ | - |
| CVaR (Martin et al., 2003): $E[r_p]/E[-r_p \mid r_p \leq -\text{VaR}_\alpha(r_p)]$ | $\alpha \in (0,1)$ |
| Rachev (Biglova et al., 2004): $E[r_p \mid r_p \geq \text{VaR}_\beta(r_p)]/E[-r_p \mid r_p \leq -\text{VaR}_\alpha(r_p)]$ | $\alpha, \beta \in (0,1)$ |

*Note*: *Rewards for all ratios are equal to $E[r_p]$ except in Rachev. In Rachev, reward is a measure called Expected Tailed Loss (ETL) which is calculated as $ETL_{\alpha\%}(X) = E[L\mid L \geq \text{VaR}_\alpha]$ where $L = -X$.
**Also, risk measures in CVaR and Rachev are defined by ETL for some values of the risk parameter, $\alpha\%$.

## Appendix B: Proof of Proposition 1

Let $\psi = f$, using the assumption of quasi-concavity of $f$ we conclude that:

$$\nabla_x f(\mathbf{x})'(\mathbf{x}^* - \mathbf{x}) \geq 0 \qquad (\text{B}.1)$$

Now suppose, in contrary to the optimality of the point $\mathbf{x}^*$, there exists a point $\mathbf{x}$ which satisfies the constraint and $f(\mathbf{x}) > f(\mathbf{x}^*)$. So, we can pick $\epsilon > 0$ and sufficiently small to find a new point $y = x - \epsilon\nabla_x f(\mathbf{x}^*) - \mathbf{x}^*$ such that $f(y) > f(\mathbf{x}^*)$. Thus:

$$\begin{aligned}\nabla_x f(\mathbf{x}^*)(x - \epsilon\nabla_x \quad f(\mathbf{x}^*) - \mathbf{x}^*) \\ = \nabla_x f(\mathbf{x}^*)(x - \mathbf{x}^*) - \epsilon\lVert\nabla_x f(\mathbf{x}^*)\rVert^2 < 0\end{aligned} \qquad (\text{B}.2)$$

Given the assumption that $\nabla_x f(\mathbf{x}^*) \neq 0$, the above inequality implies that $f(x - \epsilon\nabla_x f(\mathbf{x}^*)) < f(\mathbf{x}^*)$, contradicting $f(\mathbf{x} - \epsilon\nabla_x f(\mathbf{x}^*)) > f(\mathbf{x}^*)$. Thus, in fact, $f(\mathbf{x}) < f(\mathbf{x}^*)$.

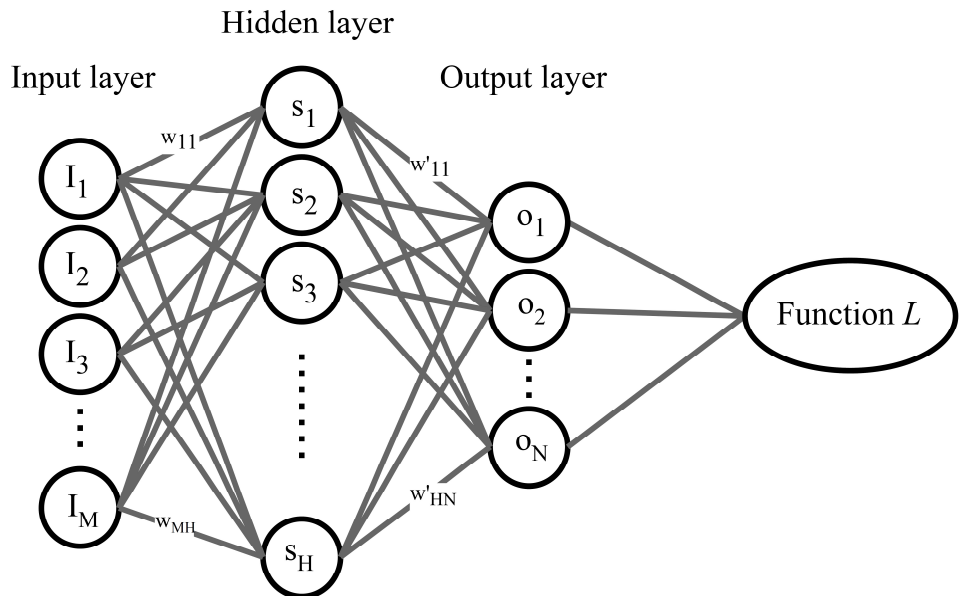

**Figure 1**: Structure of the Artificial Neural Network (ANN).

This figure shows a general structure of Artificial Neural Network (ANN). It includes input, hidden, and output layers with some functions in the hidden layer. Inputs are multiplied by weights (edges) and plugged into nodes in the hidden layer. Nodes have functions (activation functions) that generate outputs. The outputs are used in a loss function with the aim to minimize the loss by optimal changes in the weights (edges) of the network. The predictive power can be improved by increasing the number of hidden layers but there is a risk of overfitting (capturing noise rather information) if they increase excessively.

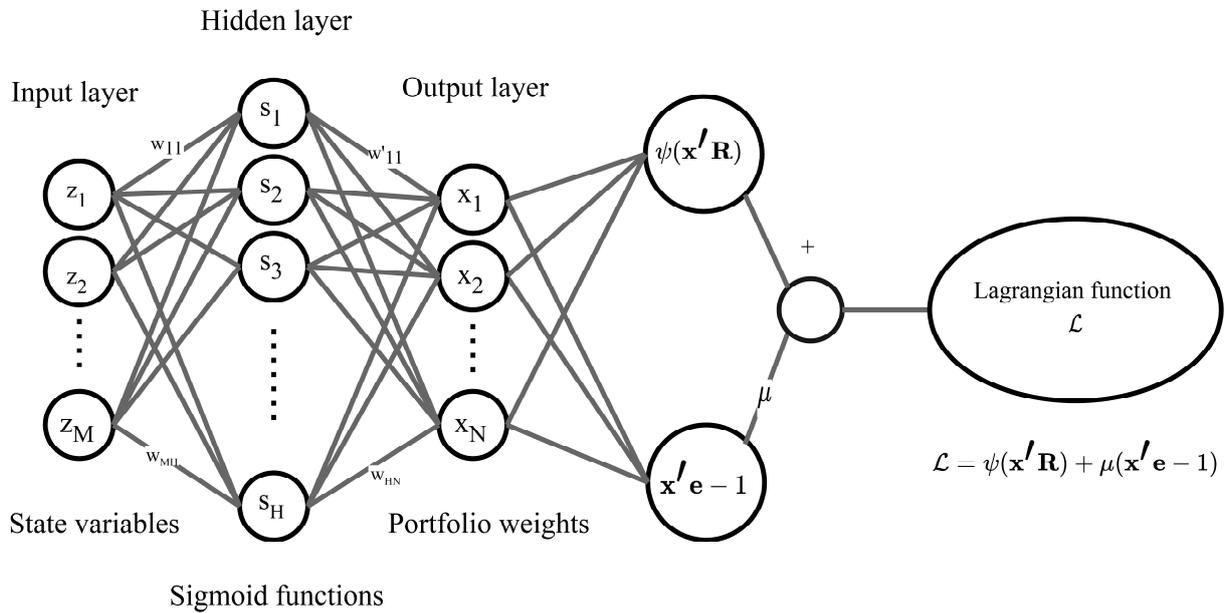

**Figure 2**: Structure of the proposed ANN to estimate portfolio choice.
This figure shows the structure of ANN designed to determine a model for portfolio choice. The input includes state variables. Sigmoid functions are used in hidden nodes. The output is portfolio choice. In order to construct the Lagrangian function, the output of the network is incorporated into both the performance ratio $\psi$ and the constraint $\mathbf{x}'\mathbf{e} - 1$. The constraint is multiplied by the Lagrangian multiplier $\mu$ and added to the ratio.

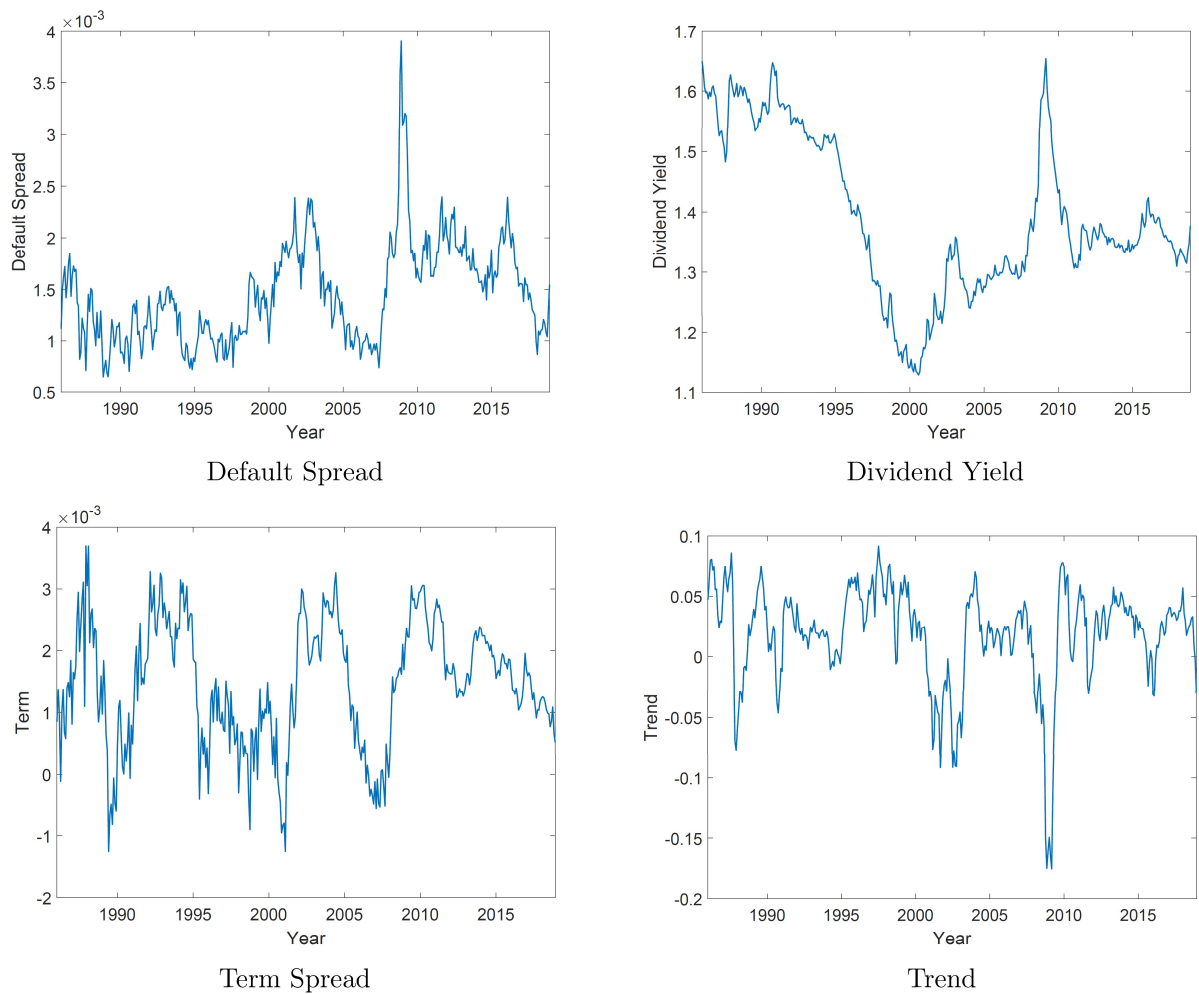

**Figure 3**: Time series of state variables.

This figure shows monthly time series plots of four state variables; Default Spread, Dividend Yield, Term Spread and Trend. Default Spread is the monthly difference between Moody's Baa- and Aaa-rated Corporate Bond Yields, Dividend Yield is log dividend-to-price ratio, Term Spread is the monthly difference between the rates on 10- and 1-year Treasury Constant Maturity Rates, and Trend is the log of the ratio of the current S&P 500 Index level and the average Index level over the previous 12 months. The sample is from January 1986 to December 2018.

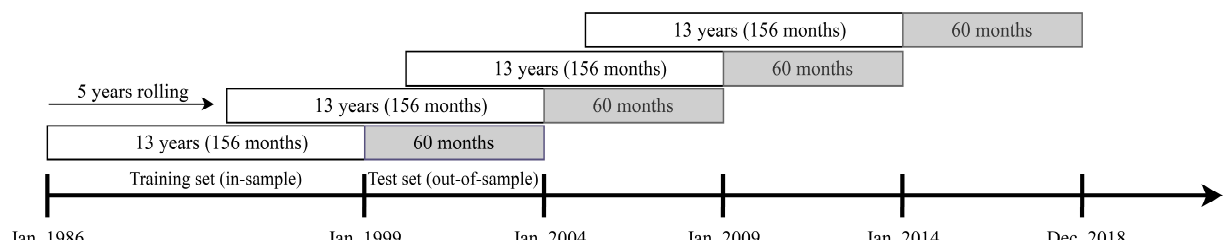

**Figure 4**: Sample period partitioning and rolling windows.
This figure shows how the entire sample period of January 1986 to December 2018 is partitioned to training (in-sample) and test (out-of-sample) samples. We use 13 years of data starting from January 1986 as the training set and the following five years as the out-of-sample set. We rollover these sets for five years until the end of the sample in December 2018. This approach generates four training sets and four consecutive out-of-sample sets of 1999–2003, 2004–2008, 2009–2013 and 2014–2018.

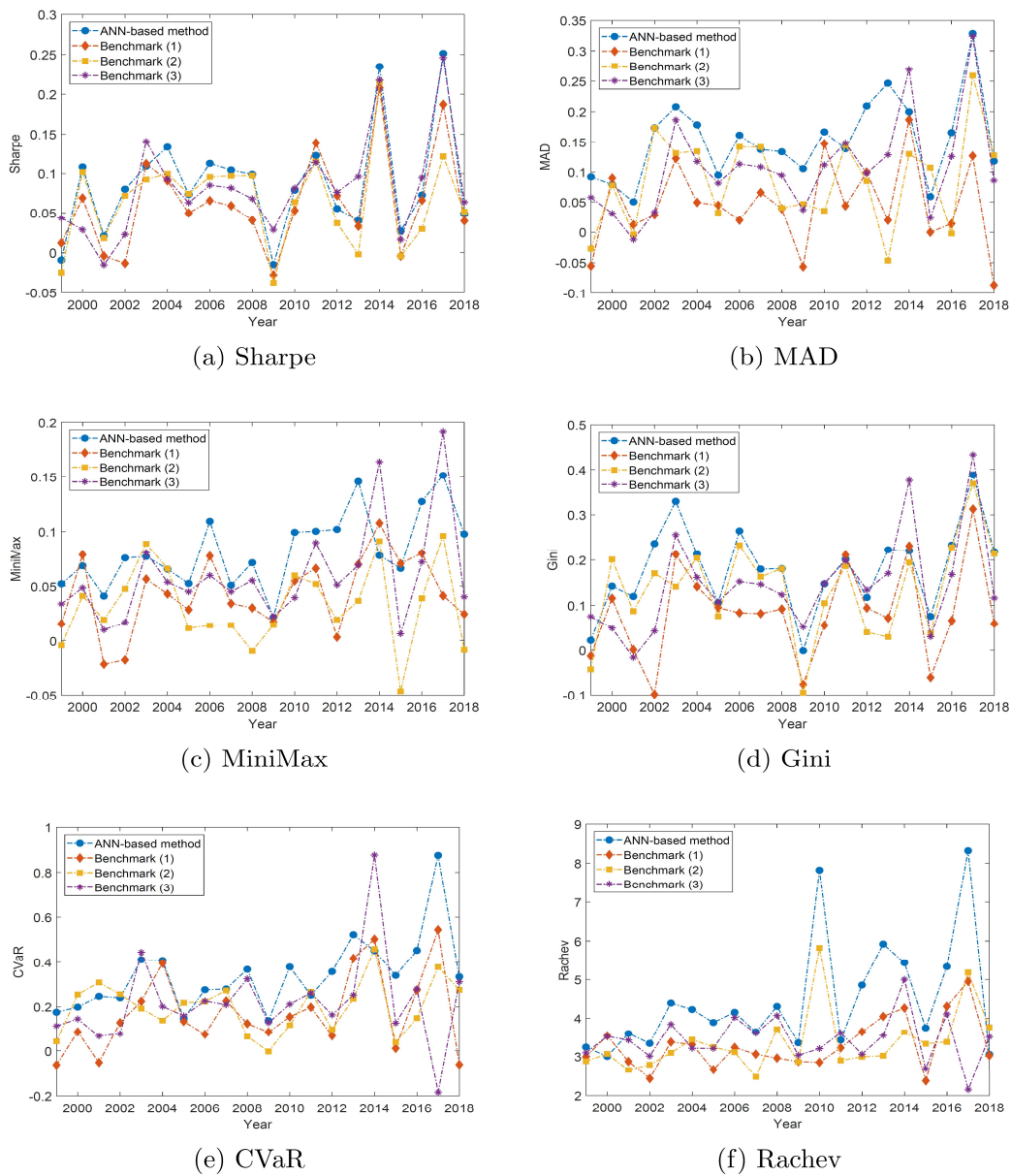

(a) Sharpe       (b) MAD

(c) MiniMax       (d) Gini

(e) CVaR       (f) Rachev

**Figure 5**: ANN-based versus traditional estimates of portfolio choice.

This figure displays the time series of performance ratios using the proposed ANN-based approach as well as three traditional methods as benchmarks. In benchmark method (1), moments of returns are estimated using an AR (1) process before maximizing the performance ratios, whereas in benchmark method (2), a four-factor model used to estimate the moments. The factors include Default Spread, Term Spread, Dividend Yield and Trend. Benchmark (3) is based on the parametric approach proposed by Brandt et. al. (2009) in which portfolio weights are directly estimated using state variables and by maximizing a CRRA utility function with risk parameter of $\gamma = 5$. Details of benchmark methods are explained in the text. These approaches are used for the four out-of-sample periods (1999–2003, 2004–2008, 2009–2013 and 2014–2018) to estimate optimal weights and compute the value of the ratios for each month. For each out-of-sample period and for each ratio, the prior 13 years' data are used to train ANN and estimate the parameters in benchmark methods. The plots show the monthly average of the value of performance ratios over each year from 1999 to 2018.

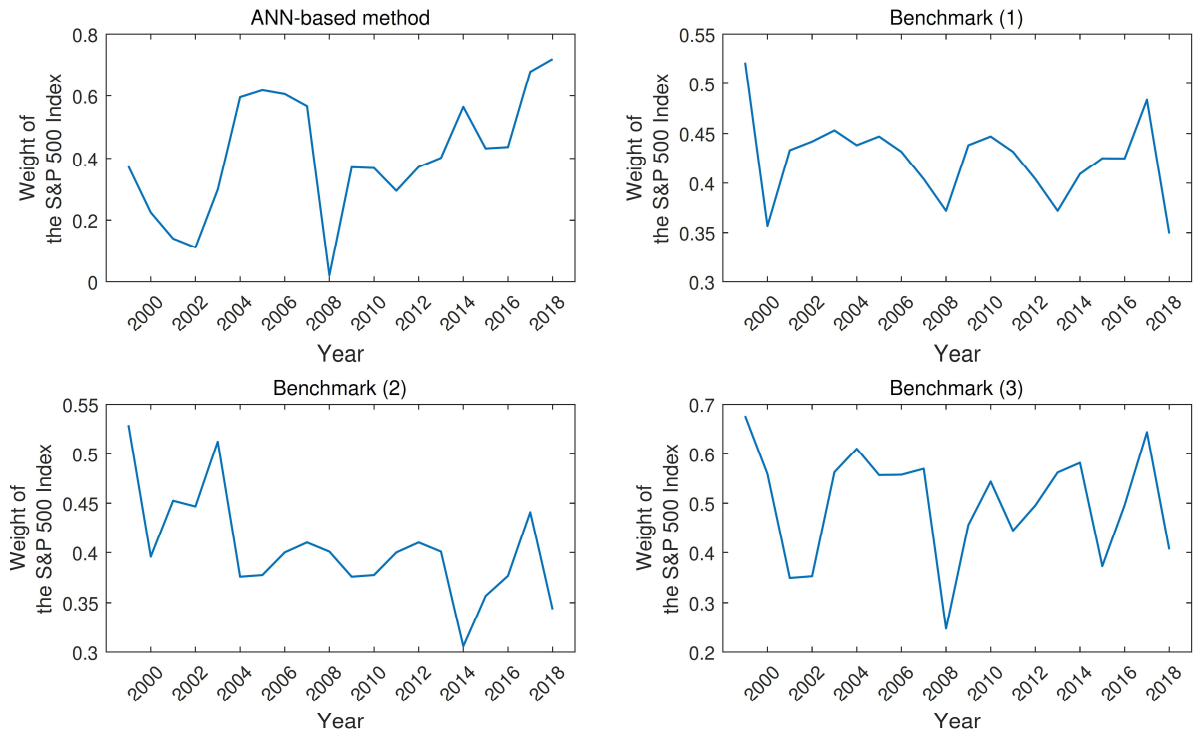

**Figure 6:** Optimal portfolio weights.
Each plot shows the average of the estimated optimal weights of the S&P 500 Index in portfolios in each year using the proposed ANN-based approach as well as three traditional methods as benchmarks. The weight of the Bond Index is equal to 1- the weight of the S&P 500 Index. In benchmark method (1), moments of returns are estimated using an AR (1) process before maximizing the Sharpe ratio, whereas in benchmark method (2), a four-factor model used to estimate the moments. The factors include Default Spread, Term Spread, Dividend Yield and Trend. Benchmark (3) is based on the parametric approach proposed by Brandt et. al. (2009) in which portfolio weights are directly estimated using state variables and by maximizing a CRRA utility function with risk parameter of $\gamma = 5$. Details of benchmark methods are explained in the text. These approaches are used for the four out-of-sample periods (1999–2003, 2004–2008, 2009–2013 and 2014–2018) to estimate optimal portfolio weights and compute the value of the ratios for each month. For each out-of-sample period and for each ratio, the prior 13 years' data are used to train ANN and estimate the parameters in benchmark methods. The plots show the average of the monthly estimated optimal weights of the S&P 500 Index over each year from 1999 to 2018.

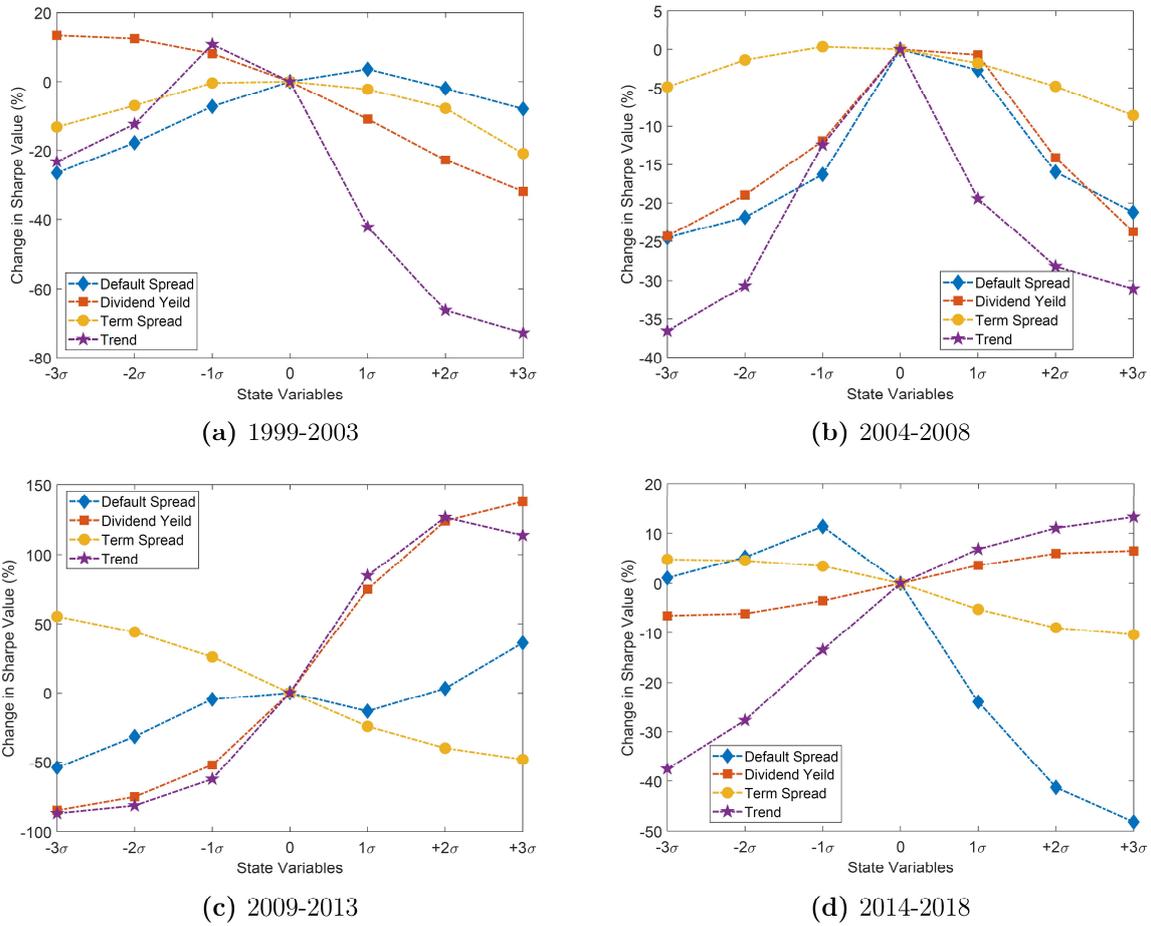

**Figure 7**: Sensitivity analysis for state variables against Sharpe ratio in the ANN network.
Panels (a) to (d) display the sensitivity of Sharpe ratio to changes in different state variables across out-of-sample periods of 1999–2003, 2004–2008, 2009–2013 and 2014–2018, respectively. For each out-of-sample period, the prior 13 years' data are employed to train the ANN for that sample period using Sharpe ratio as the performance ratio. We use the ANN and the values of state variables, as the input of the ANN, at the beginning of each month of the out-of-sample period to estimate the optimal portfolio choice and calculate the monthly average of the Sharpe ratio. The state variables are demeaned and standardized. We change the values of one state variable across a fixed range, while the values of other state variables are actual data. The range of change is $\pm 3\sigma$, $\pm 2\sigma$, and $\pm \sigma$, where $\sigma$ is the standard deviation of the state variable. We compute the ranges of changes for Sharpe ratio against the changes in values of the state variable. Each curve in each panel shows the sensitivity analysis for the associated state variable.

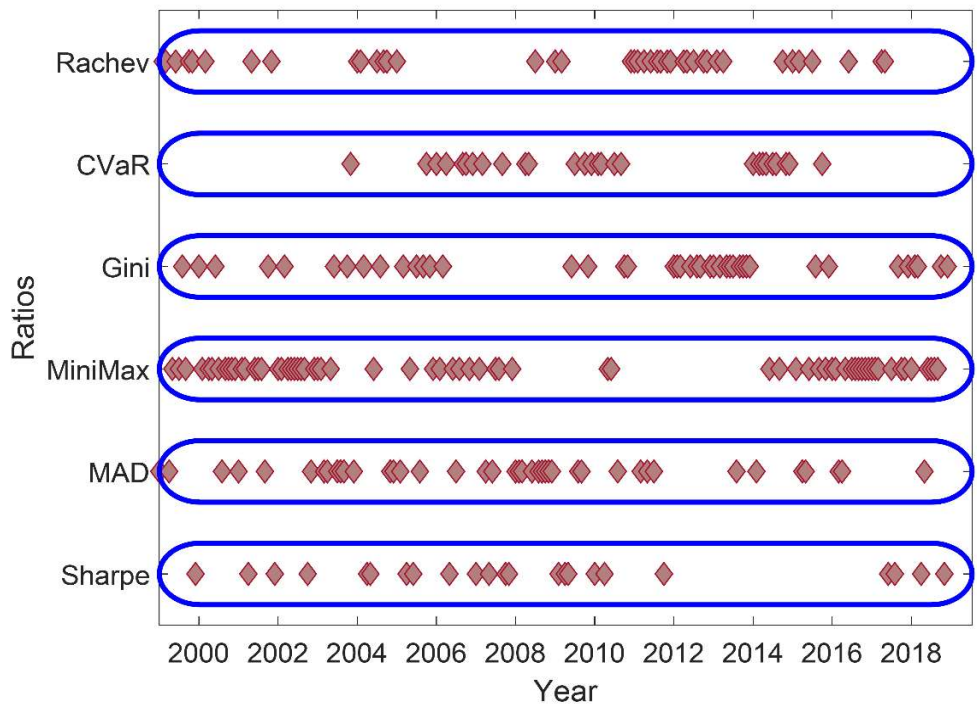

**Figure 8**: Best performed ratio over time.
This figure shows the performance ratio that generates the highest wealth in each month. For each of the out-of-sample periods of 1999–2003, 2004–2008, 2009–2013 and 2014–2018, we use the prior 13 years' monthly data to estimate the ANN (training ANN) of the portfolio choice for each performance ratio. Then we use the state variables at the beginning of each month (of the out-of-samples) and the trained ANN for each ratio and estimate the optimal portfolio weights and monthly portfolio returns. In each month, we compare the returns and identify the performance ratio that resulted in the highest monthly return. The figure displays how this ratio (so-called 'best ratio') is distributed across different ratios over time.

**Table 1**: Descriptive statistics of returns and state variables.
This table presents annualized descriptive statistics of daily returns for the S&P 500 Index and the Bond Index. The table also shows descriptive statistics of the four state variables as follows. The Default Spread is the monthly difference between Moody's Baa- and Aaa-rated Corporate Bond Yields, the Dividend Yield is the log dividend-to-price ratio of the S&P 500 Index, the Term Spread is the monthly difference between the rates on 10- and 1-year Treasury Constant Maturity Rates, and the S&P 500 Index momentum variable Trend is the log of the ratio of the current S&P 500 Index level and the average Index level over the previous 12 months.

|                | Mean | Median | StdDev | Skewness | Kurtosis |
|----------------|------|--------|--------|----------|----------|
| S&P            | 0.09 | 0.08   | 0.18   | -0.84    | 24.94    |
| Bond           | 0.08 | 0.06   | 0.13   | -0.03    | 6.17     |
| Default Spread | 0.15 | 0.14   | 0.05   | 1.06     | 5.42     |
| Dividend Yield | 1.39 | 1.36   | 0.13   | 0.16     | 2.06     |
| Term Spread    | 0.15 | 0.15   | 0.10   | -0.30    | 2.52     |
| Trend          | 1.81 | 2.49   | 4.18   | -1.65    | 7.27     |

**Table 2**: ANN-based versus traditional estimates of portfolio choice.
This table shows the descriptive statistics of the performance ratios maximized using the proposed ANN-based as well as by three benchmark methods. In benchmark method (1), moments of returns are estimated using an AR (1) process before maximizing the performance ratios, whereas in benchmark method (2), a four-factor model used to estimate the moments. The factors include Default Spread, Term Spread, Dividend Yield and Trend. Benchmark (3) is based on the parametric approach proposed by Brandt et. al. (2009) in which portfolio weights are directly estimated using state variables and by maximizing a CRRA utility function with risk parameter of $\gamma = 5$. Details of benchmark methods are explained in the text. These methods are used for the four out-of-sample periods (1999–2003, 2004–2008, 2009–2013 and 2014–2018) to estimate optimal portfolio weights and compute the ratios for each month. For each out-of-sample period and for each ratio, the prior 13 years' data are used to train ANN and estimate the parameters of the benchmark models. The table presents the summary statistics of monthly value of performance ratios over the whole period of 1999–2018. Performance ratios are Sharpe, MAD, MiniMax, Gini, CVaR and Rachev. The symbols ∗∗∗, ∗∗, and ∗ indicate that the mean difference between ANN-based approach and the best benchmark is significantly different from zero at the 1%, 5%, and 10% levels, respectively.

| Methods | Mean | StdDev | Skewness | Kurtosis |
|---|---|---|---|---|
| Panel A: Sharpe ratio | | | | |
| ANN-based method | 0.09 | 0.20 | 0.12 | 3.28 |
| Benchmark (1) | 0.06 | 0.20 | -0.05 | 2.88 |
| Benchmark (2) | 0.07 | 0.20 | -0.10 | 2.73 |
| Benchmark (3) | 0.08 | 0.20 | 0.38 | 3.77 |
| Panel B: MAD ratio | | | | |
| ANN-based method | 0.15*** | 0.26 | 0.06 | 3.35 |
| Benchmark (1) | 0.05 | 0.24 | 0.07 | 2.80 |
| Benchmark (2) | 0.09 | 0.26 | 0.05 | 2.95 |
| Benchmark (3) | 0.10 | 0.25 | 0.35 | 3.66 |
| Panel C: MiniMax ratio | | | | |
| ANN-based method | 0.08*** | 0.14 | 1.43 | 6.32 |
| Benchmark (1) | 0.04 | 0.12 | 1.27 | 7.58 |
| Benchmark (2) | 0.03 | 0.12 | 0.91 | 5.22 |
| Benchmark (3) | 0.06 | 0.14 | 2.98 | 22.42 |
| Panel D: Gini ratio | | | | |
| ANN-based method | 0.18*** | 0.34 | -0.01 | 3.43 |
| Benchmark (1) | 0.08 | 0.35 | 0.08 | 2.74 |
| Benchmark (2) | 0.14 | 0.35 | 0.01 | 2.79 |
| Benchmark (3) | 0.14 | 0.35 | 0.35 | 3.61 |
| Panel E: CVaR ratio | | | | |
| ANN-based method | 0.34** | 0.64 | 3.48 | 13.26 |
| Benchmark (1) | 0.17 | 0.51 | 3.22 | 9.24 |
| Benchmark (2) | 0.20 | 0.47 | 1.98 | 8.13 |
| Benchmark (3) | 0.22 | 0.70 | -2.47 | 48.86 |
| Panel F: Rachev ratio | | | | |
| ANN-based method | 4.45*** | 4.51 | 5.26 | 40.01 |
| Benchmark (1) | 3.33 | 2.01 | 2.59 | 14.48 |
| Benchmark (2) | 3.36 | 3.10 | 8.32 | 97.88 |
| Benchmark (3) | 3.44 | 2.69 | -1.25 | 31.07 |

**Table 3**: ANN-based versus traditional estimates of portfolio choice in different subsamples.

This table shows the mean value of the performance ratios maximized during the out-of-sample periods of 1999-2003, 2004-2008, 2009-2013 and 2014-2018 using ANN as well as three benchmark methods. In benchmark method (1), moments of returns are estimated using an AR (1) process before maximizing the performance ratios, whereas in benchmark method (2), a four-factor model used to estimate the moments. The factors include Default Spread, Term Spread, Dividend Yield and Trend. Benchmark (3) is based on the parametric approach proposed by Brandt et. al. (2009) in which portfolio weights are directly estimated using state variables and by maximizing a CRRA utility function with risk parameter of $\gamma = 5$. Details of benchmark methods are explained in the text. These methods are used for the four out-of-sample periods to estimate optimal portfolio weights and compute the ratios for each month. For each out-of-sample period and for each ratio, the prior 13 years data are used to train ANN and estimate the parameters of the benchmark methods. Performance ratios are Sharpe, MAD, MiniMax, Gini, CVaR and Rachev. The symbols ∗∗∗, ∗∗, and ∗ indicate that the mean difference between ANN and the best benchmark is significantly different from zero at the 1%, 5%, and 10% levels, respectively.

| Methods | 1999-2003 | 2004-2008 | 2009-2013 | 2014-2018 |
|---|---|---|---|---|
| Panel A: Sharpe ratio | | | | |
| ANN-based method | 0.06*** | 0.10 | 0.06 | 0.13 |
| Benchmark (1) | 0.04 | 0.06 | 0.05 | 0.10 |
| Benchmark (2) | 0.05 | 0.09 | 0.04 | 0.08 |
| Benchmark (3) | 0.04 | 0.08 | 0.08 | 0.12 |
| Panel B: MAD ratio | | | | |
| ANN-based method | 0.12* | 0.14* | 0.17** | 0.17 |
| Benchmark (1) | 0.04 | 0.04 | 0.05 | 0.05 |
| Benchmark (2) | 0.07 | 0.10 | 0.06 | 0.12 |
| Benchmark (3) | 0.06 | 0.10 | 0.10 | 0.16 |
| Panel C: MiniMax ratio | | | | |
| ANN-based method | 0.06*** | 0.07* | 0.09** | 0.10 |
| Benchmark (1) | 0.02 | 0.04 | 0.04 | 0.06 |
| Benchmark (2) | 0.04 | 0.02 | 0.04 | 0.03 |
| Benchmark (3) | 0.04 | 0.05 | 0.05 | 0.09 |
| Panel D: Gini ratio | | | | |
| ANN-based method | 0.17* | 0.19 | 0.14 | 0.23 |
| Benchmark (1) | 0.04 | 0.10 | 0.07 | 0.12 |
| Benchmark (2) | 0.11 | 0.17 | 0.05 | 0.21 |
| Benchmark (3) | 0.08 | 0.13 | 0.14 | 0.22 |
| Panel E: CVaR ratio | | | | |
| ANN-based method | 0.25 | 0.30** | 0.33** | 0.49 |
| Benchmark (1) | 0.06 | 0.19 | 0.18 | 0.25 |
| Benchmark (2) | 0.21 | 0.18 | 0.14 | 0.26 |
| Benchmark (3) | 0.17 | 0.22 | 0.20 | 0.28 |
| Panel F: Rachev ratio | | | | |
| ANN-based method | 3.51 | 4.04* | 5.08*** | 5.18** |
| Benchmark (1) | 3.04 | 3.04 | 3.32 | 3.78 |
| Benchmark (2) | 2.89 | 3.19 | 3.52 | 3.86 |
| Benchmark (3) | 3.37 | 3.62 | 3.29 | 3.49 |

**Table 4**: Relative importance of the state variables as determinants of the Sharpe ratio.
This table reports the relative importance (RI) of each state variable in determining Sharpe ratio over the four out-of-sample periods (1999–2003, 2004–2008, 2009–2013, and 2014–2018). In Panel A, the 'Connection Weights' method suggested by Olden and Jackson (2002) is used to calculate RIs. In this method, final values of the network weights at the end of the training step are used to calculate RIs for each state variable. In Panel B, the 'Permutation Importance' suggested by Breiman (2001) is used to calculate RIs. In this method the values show the changes in the final average Sharpe values using the trained network and shuffling the corresponding state variable in out-of-sample periods. Average is the mean of all RIs computed for all four out-of-sample periods. The values in Panel B are in percentage. Details of these methods are explained in the text.

| State variable | Average (1999–2018) | 1999–2003 | 2004–2008 | 2009–2013 | 2014–2018 |
|---|---|---|---|---|---|
| Panel A: Connection Weight Method | | | | | |
| Default Spread | 3.4 | 1.6 | 4.6 | 2.3 | 4.9 |
| Dividend Yield | 2.3 | 1.0 | 4.6 | 2.9 | 0.5 |
| Term Spread | 1.0 | 1.5 | 0.7 | 0.3 | 1.3 |
| Trend | 5.3 | 5.9 | 8.4 | 5.1 | 1.8 |
| Panel B: Permutation Importance Method | | | | | |
| Default Spread | 0.96 | 0.52 | 1.32 | 0.23 | 1.77 |
| Dividend Yield | 0.22 | 0.04 | 0.44 | 0.35 | 0.03 |
| Term Spread | 0.15 | 0.12 | 0.15 | 0.22 | 0.11 |
| Trend | 1.54 | 3.14 | 1.65 | 0.35 | 1.03 |

**Table 5**: Ranking of performance ratios.

The ANN-based method is used for the four out-of-sample periods (1999–2003, 2004–2008, 2009–2013 and 2014–2018) to determine optimal portfolio weights and compute the portfolio return for each month. Performance ratios, as the objective function in the optimization problem, are Sharpe, MAD, MiniMax, Gini, CVaR and Rachev. For each out-of-sample period and for each ratio, the prior 13 years' data are used to train ANN. The table presents the monthly average optimal portfolio returns, associated with each performance ratio, over the whole out-of-sample period of 1999–2018. The ratios are ranked based on the monthly average returns in panel (a). We also report the number of months over the out-of-sample periods (frequency) that a performance ratio has the highest monthly returns in panel (b). Ratios in panel (b) are ranked based on the frequency with which that they have generated the highest returns during the whole out-of-sample period. The table also reports the Pearson correlation and its $p$-value (in parenthesis) between two rank series (ranks based on returns (panel (a)) and frequency (panel (b))).

| Rank | Panel (a): Monthly Return | | Panel (b): Frequency | |
|---|---|---|---|---|
| | Ratio | Value | Ratio | Value |
| 1 | Rachev | 1.08 | MiniMax | 67 |
| 2 | MiniMax | 1.02 | Gini | 42 |
| 3 | CVaR | 0.95 | MAD | 41 |
| 4 | MAD | 0.89 | Rachev | 40 |
| 5 | Gini | 0.85 | CVaR | 27 |
| 6 | Sharpe | 0.85 | Sharpe | 23 |
| Correlation | 0.31 (0.54) | | The whole out-of-sample period= 240 months | |